\documentclass[a4paper,12pt]{article}

\begin{document}
\newcommand{\be}{\begin{equation}}
\newcommand{\ee}{\end{equation}}
\newcommand{\ba}{\begin{array}}
\newcommand{\ea}{\end{array}}
\newcommand{\bea}{\begin{eqnarray}}
\newcommand{\eea}{\end{eqnarray}}
\newcommand{\bi}{\begin{itemize}}
\newcommand{\ei}{\end{itemize}}

\newcommand{\x}{{\ensuremath{\times}}}
\newcommand{\bb}[1]{\makebox[16pt]{{\bf#1}}}

\title{Completeness of superintegrability in two-dimensional constant
curvature spaces}
\author{E.\ G.\ Kalnins and J.\ M.\ Kress\\
{\sl Department of Mathematics, University of Waikato,}\\
{\sl Hamilton, New Zealand,}\\
{\sl e.kalnins@waikato.ac.nz and jonathan@math.waikato.ac.nz}\\ \\
G.\ S.\ Pogosyan\\
{\sl Centro de Ciencias Fisicas, Universidad Nacional Autonoma
de Mexico,}\\
{\sl Apartado Postal 48--3, 62251 Cuernavaca, Morelos, Mexico} \\
{\sl and} \\
{\sl Laboratory of Theoretical Physics, Joint Institute for Nuclear
 Research,}\\
{\sl Dubna, Moscow Region, 14980, Russia,}\\
{\sl pogosyan@thsun1.jinr.dubna.su}\\ \\
and W.\ Miller, Jr. \\
{\sl School of Mathematics, University of Minnesota,}\\
{\sl Minneapolis, Minnesota, 55455, U.S.A.,}\\
{\sl miller@ima.umn.edu}}
\date{\today}
\maketitle

\begin{abstract}
We classify  the Hamiltonians $H=p_x^2+p_y^2+V(x,y)$ of
all classical superintegrable systems in two dimensional complex Euclidean space with
second-order  constants of the motion. We similarly classify the
superintegrable Hamiltonians 
$H=J_1^2+J_2^2+J_3^2+ V(x,y,z)$ on the complex 2-sphere where
$x^2+y^2+z^2=1$. This is 
 achieved in all generality using properties of the complex Euclidean
group and the complex orthogonal group.
\end{abstract}

\section{ Introduction}
It is known from classical mechanics that a mechanical system with $n$
degrees of freedom is completely 
integrable if there are $n$ functionally independent constants of the
motion which are 
mutually in involution \cite{VIA}. The idea of a superintegrable system 
is that there 
exist more than $n$ functionally independent constants of the motion,
but not necessarily in involution. If there are $2n-1$ such 
constants the system is said to be {\it maximally superintegrable} or 
just {\it superintegrable} \cite{EVA,EVAN,WOJ,RAN}. 
Here we consider only the case where there
exist $2n-1$ functionally independent constants of the motion
(including the Hamiltonian) that are quadratic in the momenta.
Ra\~nada \cite{RAN} investigated such systems and noted that many
could be found in Drach's list of potentials admitting 
constants cubic in the momenta.\footnote{These are
{\bf [E1,E2,E7,E9,E16,E19,E20]} in out notation.}
In the papers \cite{KMJP2,KMJP3,KMJP4} we have given a complete 
classification of all
{\it non-degenerate potentials} on complex Euclidean 2-space and on the
complex 2-sphere that give rise to superintegrable systems. (For
example in \cite{KMJP2} we have calculated  all
the inequivalent superintegrable potentials $V$ that are
{\sl non-degenerate} in the sense that they   depend uniquely
on four arbitrary parameters, i.e., one can prescribe the values of
$V,V_x,V_y,V_{yy}$ arbitrarily at any regular point $(x_0,y_0)$ and
these values determine $V(x,y)$ uniquely.) In 
this article we relax this  requirement   and ask the same question but 
without 
the condition of non-degeneracy: For which potentials in two dimensions 
do there exist at least two  constants of the motion 
\be
A_j=a_j(x,y)p^2_x+b_j(x,y)p^2_y+c_j(x,y)p_xp_y+d_j(x,y)=A'_j+d_j\,,
\quad 
j=1,2
\label{com}
\ee
in addition to the Euclidean space Hamiltonian
\be
H=p^2_x+p^2_y+V(x,y),\label{Ham}
\ee
i.e., $\{H,A_j\}=0,\quad j=1,2$, and such that the $2n-1=3$ constants
of the motion $H=A_0,A_1,A_2$ are functionally independent on phase
space? We will do the same for the Hamiltonian on the complex 2-sphere
\be
H=J_1^2+J_2^2+J_3^2+V(x,y,z),\label{Ham1}
\ee
where $x^2+y^2+z^2=1$ and $J_1=yp_z-zp_y$, $J_2=zp_x-xp_z$,
$J_3=xp_y-yp_x$.
We give a complete solution. The computations are lengthy, and
throughout we have made use of a computer algebra package. We give
many details in the first few examples, to make our method clear.

In references \cite{KMJP2,KMJP3,KMJP4}
we worked out the structure of the quadratic algebra for each of the
non-degenerate potentials. In this paper we supplement those results by
computing the quadratic algebras for the non-constant degenerate
potentials. Also we correct a few errors and fill in some gaps in
those earlier papers.

\section{Superintegrability in $E_{2,C}$}
In the computations to follow, quite often we will be considering
systems such that the corresponding Hamilton Jacobi equation 
$H=E$, (\ref{Ham}), can be solved by the 
method of separation of variables. Then
$p_x=\frac{\partial S}{\partial x} ,p_y=\frac{\partial S}{\partial y}$
and there is a complete integral of the form  
$$
S=U(u,E,\lambda)+V(v,E,\lambda)
$$
for separable coordinates $u=u(x,y),v=v(x,y)$ and some separation 
constant $\lambda$. This will not always be the case, but when
separation is possible a knowledge of the separable coordinates ${u,v}$
will greatly simplify our computations. In general we will use the
structure of the complex Euclidean group $E(2,C)$ and its Lie algebra
$e(2,C)$ to solve our problem. The elements  $L_1=p_x,\ L_2=p_y$ and 
$L_3=M=xp_y-yp_x$ form a basis for $e(2,C)$ under the Poisson bracket,
and quadratic elements ${L_iL_j, \quad i\le j}$ form a basis for all
purely quadratic functions $A'$ such that $\{A',p_x^2+p_y^2\}=0$. Thus
the quadratic integrals $A_j$ can be written in the form  
\be
A_j=a_j^{k\ell}L_kL_\ell+d_j(x,y)=A'_j+d_j
\ee
for suitable constants $a_j^{k\ell}=a_j^{\ell k}$.  

We will not regard Euclidean space Hamiltonians as essentially 
different if they are related by a Euclidean transformation. Because of 
this we can use Euclidean group transformations to simplify the 
expressions for the $A_j$ and classify them into equivalence classes. 
If we do this then there are   
equivalence classes of constants whose typical representatives are \cite{MIL}
\be
p^2_x ,\quad  (p_x+ip_y)^2 ,\quad M^2 ,\quad
M(p_x+ip_y)+(p_x-ip_y)^2,\label{equivclass}
\ee
$$
M^2+(p_x+ip_y)^2,\quad Mp_x ,\quad M^2+c^2p^2_x ,\quad  M(p_x+ip_y).
$$
(Note that, up to the addition of an arbitrary  multiple of the Casimir
element $p_x^2+p_y^2$, this is simply a choice of a representative on each
distinct orbit of second order elements in the enveloping algebra of
$e(2,C)$ under the adjoint action of $E(2,C)$.)
Without loss of generality, we can assume that $A'_1$ 
coincides with one of these representatives and use the defining
relations $\{H,A_j\}=0$ for constants of the motion,
to determine the general form of the second constant. This is a 
particularly useful strategy since all but the last of the  
list (\ref{equivclass}) of representatives has a form that implies 
separation of variables in at 
least one coordinate system. If we use this fact then we see that 
the corresponding 
potential $V$ must have the form implied by separation, and the 
requirement of an extra 
constant of the motion implies strong conditions on this functional
form.  This will greatly simplify our computations. 
For all but one potential, we find that
the associated constants determine more than
one separating coordinate system.  These are listed in Appendix
\ref{appendix1}.

We now deal with each of these cases individually.  Consider the first
constant in our complete family of equivalence classes. We assume
our Hamiltonian has a constant of the form   
\be
A_1=p^2_x+d_1(x,y).\label{px}
\ee
The condition $\{H,A_1\}=0$ implies 
$\partial_yd_1(x,y)=0,\quad \partial_x(V-d_1(x,y))=0$,
so we can assume that the Hamiltonian has the form 
\be
H=A_0=p^2_x+p^2_y +f(x)+h(y)
\ee
where $d_1(x,y)=f(x)$. For superintegrability we must have one 
additional constant of the motion which can be written  
\be
A_2=a_2^{k\ell}L_kL_\ell+d_2(x,y).
\ee
Since we can always add linear combinations of $A_0$ and $A_1$ to
$A_2$ without changing the system, we can assume that 
\be
A_2=AM^2+BMp_x+CMp_y+Dp_xp_y+d_2(x,y)\label{case1}
\ee
where $A,B,C,D$ are constants, not all zero. 

Because of the forms of $A_0$ and 
$A_1$ we can always apply translations to $A_2$ in order to simplify its 
form. Suppose $A\ne 0$ in (\ref{case1}). Then, normalizing so that $A=1$, by appropriate
translations in $x$ and $y$ we can pass to a new Cartesian coordinate
system in which $B=C=0$. 
The condition   
$\{A_0,A_2\}=0$
then determines the possible forms of $A_2$. Indeed, equating
coefficients of $p_x$ and $p_y$, we find
$$
\partial_y d_2(x,y)=(D-2xy)f'(x)+2x^2h'(y),\qquad \partial_x
d_2(x,y)=2y^2f'(x)+(D-2xy)h'(y).
$$ 
Equating the cross partial derivatives of $d_2(x,y)$ we obtain the
condition
\be
(f''+\frac{3}{x}f')-(h''+\frac{3}{y}h')=\frac{D}{2xy}(h''-f'').\label{intcond}
\ee
If $D=0$, the variables separate and we find a well known 
non-degenerate superintegrable potential.

\bi
\item[{\bf [E1]}]
$\displaystyle V=\omega ^2(x^2+y^2)+ \frac\alpha{x^2} + \frac\beta{y^2}$.

\noindent
The additional constant has the form  
$$
A_2=M^2+ \alpha  {y^2\over x^2} +\beta  {x^2\over y^2}.
$$
%Taking linear combinations of $A_1$ and $A_2$ and refering to
%Appendix \ref{appendix1}, it is clear that this potential separates in polar
%and elliptic coordinates as well as Cartesian coordinates.
This example 
%is denoted $U^b$ in \cite{RAN} and
together with its Poisson bracket relations is well 
studied \cite{KMJP1,KMJP2}.
\ei

\noindent
If $A=0$ but $B^2+C^2\ne 0$ in (\ref{case1}) then we can rotate
coordinates so that $B=0$, normalize so that $C=1$, translate to obtain
$D=0$, and find the non-degenerate potential

\bi
\item[{\bf [E2]}]
$\displaystyle V=\omega ^2(4x^2+y^2)+\alpha x+{\beta \over y^2}$

\noindent
The additional constant has the form  
$$
A_2=Mp_y+d_2(x,y)\,.
$$
%which corresponds to separation in parabolic coordinates.
%This is potential $U^a$ in \cite{RAN}.
\ei

\noindent 
There are no further non-degenerate potentials separating in Cartesian 
coordinates.  We now return to the case $A\neq0$ and
suppose that $D\ne 0$. Then from  (\ref{intcond}) we see that $h'',f''$
must satisfy a functional equation of the form
$h''(y)-f''(x)=xy(G(y)+F(x))$ for some functions $G,F$. Solving this
equation and substituting back into (\ref{intcond}), we find that the
variables separate and we obtain the solution

\bi
\item[{\bf [E3]}]
$\displaystyle V=\omega ^2(x^2+y^2)$.

\noindent
In this case, $d_1(x,y)=\omega^2x^2$, and 
there are two additional constants, one of which is first order.
They can be taken in the form
$$
A_2=p_xp_y+\omega ^2xy\,, \qquad X = M\,.
$$
The Poisson bracket relations for these constants are
$$\{A_1,X\}=2A_2\,,\quad \{A_2,X\}=A_0-2A_1\,,\quad 
\{A_1,A_2\}=-2\omega ^2X\,.
$$
Since $A_0,A_1,A_2$ are functionally independent, all constants of the
motion are functions of these. It is easy to verify that $X$ 
satisfies the functional relation
$$
A^2_2-A_1(A_0-A_1)+\omega ^2X^2=0.
$$
This example is the harmonic oscillator in two dimensions.
\ei

\noindent
If $A=0$ and $B^2+C^2=0$ but $B\ne 0$, we can take $B=1$, $C=i$ (by
mapping $y$ to $-y$ if necessary), and translate to get $D=0$. A
straightforward computation gives 

\bi
\item[{\bf [E4]}]
$\displaystyle V=\alpha(x+iy)$.

\noindent
Here $d_1(x,y)=\alpha x$ and the Hamiltonian admits two extra
constants, one of which is first order,
$$
A_2=M(p_x+ip_y)+\frac{i\alpha}{4}(x+iy)^2\,, \quad X = p_x+ip_y\,.
$$
The Poisson bracket relations take the form
$$
\{A_1,X\}=-\alpha,\quad \{A_2,X\}=iX^2,\quad
\{A_1,A_2\}=-iX^3+2iA_1X-iA_0X,
$$
with the functional relation
$$
A_0^2 + X^2(2A_0-4A_1+X^2)+4i\alpha A_2=0\,.
$$
\ei

\noindent
If $A=B=C=0$ then we normalize $D=2$ and find the non-degenerate
potential  $V=\omega^2(x^2+y^2)+\alpha x +\beta y.$
For a fixed choice of the parameters, the Hamiltonian admits the first
order constant of the motion
$$
X=2\omega^2M+\alpha p_y-\beta p_x.
$$
By an appropriate translation we can obtain $X'=M$, which is case {\bf [E3]}.

There are special cases of  potentials {\bf [E1]} and {\bf [E2]} such that the
Hamiltonian admits more than two constants of the motion.  The
 possibilities are
\bi
\item[{\bf [E5]}]
$\displaystyle V=\alpha x$.

\noindent
Since in this case, $A_0-A_1=p_y^2$, we can replace $A_1$ with the first
order constant $p_y$ and take the additional constants as
$$
A_2=Mp_y-\frac\alpha4y^2\,, \quad  
A_3=p_xp_y+\frac\alpha2y\,, \quad X=p_y\,.
$$
They satisfy the Poisson bracket relations 
$$
\{A_2,X\}=A_3\,, \quad \{A_3,X\}=-\frac\alpha2\,,  \quad
\{A_3,A_2\}=2X^3-A_0X
$$
and the functional relation
$$
A_3^2+X^4-A_0X^2+\alpha A_2=0\,.
$$
\item[{\bf [E6]}]
$\displaystyle V=\frac\alpha{x^2}$.

\noindent
As for the previous case, we can replace $A_1$ with $p_y$.
The additional constants are  
$$
A_2=Mp_x-\frac{\alpha y}{x^2}\,,\quad  
A_3=M^2+ \frac{\alpha y^2}{x^2},\quad X=p_y\,.
$$
Their Poisson brackets are  
$$
\{A_2,X\}=A_0-X^2\,, \quad \{A_3,X\}=2A_2\,, \quad  
\{A_3,A_2\}=-2XA_3-2\alpha X\,,
$$
and they satisfy the functional relation  
$$
A_2^2-A_3(A_0-X^2)+\alpha X^2=0\,.
$$
\ei

\noindent
This concludes the list of possible potentials corresponding to the 
first equivalence 
class of second order elements in the enveloping algebra of $e(2,C)$.

For orbits of the second type, the constant of the motion $A_1$ has the form  
\be
A_1=p_-^2+d_1(x,y).\label{pplus}
\ee
(We adopt the notation $p_\pm=p_x\pm ip_y, z=x+iy,{\bar z}=x-iy$.)
It follows from the relation $\{A_0,A_1\}=0$ that for a Hamiltonian to 
admit a constant of the motion of  
the form (\ref{pplus}) the potential $V$ must have the form   
$$
V=f({\bar z})z+h({\bar z}),
$$
for  some functions $f$ and $h$. We can assume
$$
A_2=AM^2+BM{p_+} +CM{p_-} +Dp_+^2+d_2(z,{\bar z}).\label{zpot}
$$
There are several
possibilities. In the first case we assume $A\ne 0$. Then we can
normalize $A=1$, translate to get $B=C=0$, and write $D=c^2/2$. We
find the non-degenerate potential \cite{KMJP2}

\bi
\item[{\bf [E7]}]
$\displaystyle V = \frac{\alpha {\bar z}}{\sqrt{{\bar z}^2-c^2}}
   + \frac{\beta z}{\sqrt{{\bar z}^2-c^2}({\bar z}+\sqrt{{\bar z}^2-c^2})^2} 
   + \gamma z{\bar z}$.

\noindent
Here the second constant of the motion can be taken in the form  
$$
A_2=M^2+c^2p^2_x +d_2(x,y).
$$
\ei

\noindent
The limiting case of this as $c\to 0$ gives  the potential \cite{KMJP2}
\bi
\item[{\bf [E8]}]
$\displaystyle V=\frac{\alpha z}{{\bar z}^3} + \frac\beta{{\bar z}^2}
      + \gamma z{\bar z}$.

\noindent
Here the second constant of the motion has the form 
$$
A_2=M^2+d_2(x,y)\,.
$$
\ei

\noindent
If $A=0$ but $BC\ne 0$, we can normalize and rotate to obtain $B=-i/2,
C=i/2$, and translate to get $D=0$. We obtain the non-degenerate
potential \cite{KMJP2}

\bi
\item[{\bf [E9]}]
$\displaystyle V= \frac\alpha{\sqrt{\bar z}} + \beta x
      + \frac{\gamma(x+{\bar z})}{\sqrt{\bar z}}$.

\noindent
The second constant of the motion is 
$$
A_2=Mp_y+d_2(z,{\bar z}).
$$
\ei

\noindent
If $A=BC=0$ but $C\ne 0$, and $D\ne 0$  we normalize $C=4i$ and rotate
so that $D=1$ to obtain the non-degenerate potential \cite{KMJP2}

\bi
\item[{\bf [E10]}]
$\displaystyle V=\alpha {\bar z} + \beta (z - \frac32{\bar z}^2)
        + \gamma (z{\bar z} - \frac12{\bar z}^3)$.

\noindent
Here the second constant of the motion has the form  
$$
A_2=4iMp_-+p_+^2 +d_2(x,y).
$$
\ei

\noindent
If $A=BC=0$ but $B\ne 0$, (or if $C\ne 0,B=0$ and $D=0$ and we reflect
$y\to -y$), we can normalize $B=1$ and translate so that $D=0$ to
obtain the non-degenerate potential 

\bi
\item[{\bf [E11]}]
$\displaystyle V =\alpha z+ \frac{\beta z}{\sqrt {\bar z}}
     + \frac\gamma{\sqrt{\bar z}}$.

\noindent
Here the second constant of the motion has the form  
$$
A_2=Mp_++d_2(x,y).
$$
\ei

\noindent
There are special cases of potentials {\bf [E7,E8,E11]}  
that admit two extra constants of the motion.   In each of these cases
$A_1=p_-^2$,  i.e., $d_1(x,y)=0$ and  hence $X=p_-$ is a constant of the 
motion. The possibilities are 

\bi
\item[{\bf [E12] }]
$\displaystyle V= \frac{\alpha {\bar z}}{\sqrt{{\bar z}^2+c^2 }}$

\noindent
with the constants of motion given by  
$$
X=p_-\,,\quad 
A_2=M^2 - \frac{c^2}{4}p_+^2 - \frac{\alpha c^2 z}{2\sqrt{\bar z^2+c^2}}\,,
\quad
A_3=M{p_-} + \frac{i\alpha c^2}{2\sqrt{\bar z^2+c^2}}\,.
$$
The Poisson bracket relations are   
$$
\{X,A_2\}=2iA_3\,,\quad \{X,A_3\}=iX^2\,,\quad 
\{A_2,A_3\}=-2iXA_2,
$$
with the functional relation  
$$
A_3^2 - X^2A_2 - \frac{c^2}{4}A_0^2 + \frac{\alpha^2 c^2}4=0\,.
$$
\item[{\bf [E13]}]
$\displaystyle V= {\alpha \over \sqrt{\bar z}}$

\noindent
with the constants of motion given by  
$$
X=p_-\,,\quad 
A_2=M{p_+} + \frac{i\alpha z}{2\sqrt{\bar z}}\,, \quad  
A_3=M{p_-} + \frac{i\alpha}2 \sqrt{\bar z}\,.
$$
The Poisson bracket relations are
$$
\{X,A_2\}=iA_0\,,\quad \{X,A_3\}=iX^2\,,\quad 
\{A_2,A_3\}=-2iXA_2\,,
$$
with the functional relation  
$$
A_3A_0-X^2A_2- \frac i2\alpha^2=0\,.
$$
\item[{\bf [E14]}]
$\displaystyle V = \frac\alpha{{\bar z}^2}$

\noindent
with the constants of the motion  given by  
$$
X=p_-\,,\quad 
A_2=M{p_-} - \frac{i\alpha}{{\bar z}}\,, \quad 
A_3=M^2 + \frac{\alpha z}{{\bar z}}\,.
$$
The Poisson bracket relations are  
$$
\{X,A_2\}=iX^2\,,\quad
\{X,A_3\}=2iA_2\,,\quad 
\{A_2,A_3\}=2iXA_3\,,
$$
with the corresponding functional relation 
$$
A^2_2-A_3X^2+\alpha A_0=0\,.
$$
\item[{\bf [E15]}]
$\displaystyle V=h({\bar z})$

\noindent
where $h$ is any function of $\bar z$, not necessarily as already given 
above. A constant of the motion always exists of the form  
$$
A_2=M{p_-}+{i\over 2}\int {\bar z} {dh\over d{\bar z}} d{\bar z},
$$
in addition to the constant $X=p_-$. Indeed we might take 
$h({\bar z})=\alpha {\bar z}^2$, in which case  
$$
A_2=M{p_-}+{i\over 3}\alpha {\bar z}^3.
$$
This is an example of a potential for which separation of variables occurs in 
only one coordinate system, \cite{KMJP4}. 
\ei

\noindent
For orbits of the third type, the constant of the motion $A_1$ has the form  
\be
A_1=M^2+d_1(x,y).\label{M}
\ee
In this case $V=f(r)+\frac{h(\theta)}{r^2}$ where $r,\theta$ are polar
coordinates (see  Appendix \ref{appendix1}), and $d_1(x,y)=h(\theta)$.
We can assume that the second constant takes the general form
\be
A_2=AM{p_+}+BM{p_-}+Cp_+^2+Dp_-^2+d_2(x,y).\label{case3}
\ee
If $AB\ne 0$ then we can rotate and normalize to get $A=-B=-i/2$ and
translate to achieve $C=D=0$. This gives us the non-degenerate
potential \cite{KMJP2}

\bi
\item[{\bf [E16]}]
$\displaystyle V = \frac1{\sqrt {x^2+y^2}}
  \left( \alpha + \frac\beta{x+\sqrt{x^2+y^2}} + \frac\gamma{x-\sqrt{x^2+y^2}}
  \right)$.

\noindent
Here the extra constant of the motion has the form 
$$
A_2=Mp_y+d_2(x,y).
$$
\ei

\noindent
If $AB=0$, by letting $y\to -y$ if necessary, we can normalize so that
$A=1, B=0$ and translate to get $C=D=0$. This produces the
non-degenerate potential

\bi
\item[{\bf [E17]}]
$\displaystyle V= \frac\alpha{\sqrt{z{\bar z}}}
      +  \frac\beta{z^2} + \frac\gamma{z\sqrt {z{\bar z}}}$.

\noindent
Here the extra constant of the motion has the form 
$$
A_2=M{p_+}+d_2(x,y).
$$
\ei

\noindent
If $A=B=0$, the various possibilities have already been included under
previous cases. 

There is one special case where an extra constant of the motion
exists.  If $A_1$ is $M^2$, so that $d_1(x,y)=0$ and $M$ is a constant 
of the motion, then the only additional potential is 

\bi
\item[{\bf [E18]}]
$\displaystyle V=\frac\alpha{\sqrt{x^2+y^2}}$.

\noindent
The constants of the motion can be taken as
$$
A_2=Mp_x-{\alpha\over 2} {y\over \sqrt {x^2+y^2}}\,,\quad 
A_3=Mp_y+{\alpha\over 2} {x\over \sqrt {x^2+y^2}}\,,\quad
X=M\,.
$$
The Poisson bracket relations are   
$$
\{X,A_2\}=-A_3\,,\quad
\{X,A_3\}=A_2\,,\quad
\{A_2,A_3\}=XA_0\,,
$$
and these constants satisfy
$$
A_2^2+A_3^2-X^2A_0-\frac\alpha4=0\,.
$$
This is the well known Coulomb problem in two dimensions.
\ei

\noindent
For orbits of the fourth type, the constant of the motion $A_1$ has the form  
\be
A_1=M{p_+}+p_-^2+d_1(x,y),\label{SH}
\ee
corresponding to semi-hyperbolic coordinates (see
Appendix \ref{appendix1}).  However, the only superintegrable potentials associated with
this constant of the motion have already been considered, 
{\bf [E4,E10,E11,E14]}.

For orbits of the fifth type, the first  constant of the motion  has the form  
\be
A_1=M^2 +p_+^2+d_1(x,y),\label{H}
\ee
corresponding to   hyperbolic coordinates (see 
Appendix \ref{appendix1}). The second constant of the motion can be written in the form
\be
A_2=AM{p_+}+BM{p_-}+Cp_+^2+Dp_-^2+d_2(x,y).\label{case5}
\ee
If $AB\ne 0$ there are no cases with non-constant potential. If 
$AB=0,|A|+|B|>0$ there are two new non-degenerate cases.

\bi
\item[{\bf [E19]}]
$\displaystyle V= \frac{\alpha {\bar z}}{\sqrt{{\bar z}^2-4}} 
     + \frac\beta{\sqrt{z({\bar z}+2)}}
     + \frac\gamma{\sqrt{z({\bar z}-2)}}$

\noindent
where the additional constant is   
$$
A_2=M{p_-}+d_2(x,y).
$$
%and \cite{KMJP2}
\ei

\noindent
The remaining possibilities, have been listed earlier.

For orbits of the sixth type, the first  constant of the motion  has the form  
\be
A_1=Mp_x+d_1(x,y),\label{P}
\ee
corresponding to   parabolic coordinates (see 
Appendix \ref{appendix1}). There is only one (non-degenerate) case that is not already
listed above \cite{KMJP2}:

\bi
\item[{\bf [E20]}]
$\displaystyle V = \frac1{\sqrt {x^2+y^2}}\left( \alpha
     + \beta \sqrt{x+\sqrt{x^2+y^2}} 
     + \gamma \sqrt{x-\sqrt{x^2+y^2}} \right)$

\noindent
where the extra constant of the motion is  
$$
A_2=Mp_y+d_2(x,y).
$$
Note that although this potential only separates in parabolic 
coordinate systems, it separates 
in more than one such coordinate system and hence is multiseparable.
\ei

\noindent
For orbits of the seventh type, the first  constant of the motion  has the 
form  
\be
A_1=M^2 +c^2p_x^2+d_1(x,y),\label{E}
\ee
corresponding to elliptic coordinates (see 
Appendix \ref{appendix1}), however, all superintegrable potentials separating in
an elliptic coordinate system have already been listed.

The last orbit on our list of equivalence classes has a typical representative 
$A_1=M{p_+}+d_1(x,y)$. The second constant of the motion $A_2$ must lie on
the equivalence class of one of the eight canonical types
(\ref{equivclass}). Therefore, by an Euclidean group motion
(including reflections), and by
adding multiples of $A_0$ if necessary, we can assume that the leading
terms of $A_2$ are equal to one of the eight representatives
(\ref{equivclass}). Under this  transformation $A_1$ will be
mapped to a constant of motion of the form ${\tilde A}_1 =
M{p_\pm}+ap_\pm^2+{\tilde d}_1(x,y)$.  For seven of these representatives we
have already listed all possible potentials above. Therefore the only
new case we 
need consider is when $A_2$ transforms to ${\tilde A}_2=M{p_+}+{\tilde
d}_2(x,y)$. Since ${\tilde A}_1,{\tilde A}_2$ are functionally independent
constants of the motion, we must have either 
$\tilde A_1=Mp_-+ap_-^2+d_1(x,y)$ or $a\ne 0$.
Consequently the potential under consideration must admit a quadratic constant 
of the form $p_+^2+d_3(x,y)$ or one that can be further transformed
to $Mp_x+d_3(x,y)$. However, we have already listed all
superintegrable potentials admitting a constant of one these forms. 
Thus there are no
new potentials corresponding to this orbit.

This completes the list of possible potentials involved in our 
problem. As a consequence we see that the list of 20 potentials that we have 
calculated completely solves the problem in two dimensions of when a potential 
added to a flat space admits more than one quadratic constant of the
motion.
All other cases are equivalent to these via proper complex Euclidean
transformations and reflections.

\section{Superintegrability on the  complex two-sphere}
We can also solve the similar problem on the complex sphere. Our basic
problem is  to find the superintegrable potentials for the solution 
of the Hamilton-Jacobi equation on the complex two sphere, 
\be
H=J^2_1+J^2_2+J^2_3+V(x,y,z)=E, \label{HJS}
\ee
with $x,y,z$ subject to the constraint $x^2+y^2+z^2=1$, and 
$J_1=yp_z-zp_y, J_2=zp_x-xp_z, J_3=xp_y-yp_x$. There are  five 
inequivalent separable coordinate 
systems for the zero potential equation (\ref{HJS}) and five different
quadratic orbits, see Appendix \ref{appendix2}. Typical representatives of 
these orbit classes are  
\be
(J_1-iJ_2)^2 ,\quad J_3(J_1-iJ_2) ,\quad
(J_1+iJ_2)-J^2_3,\label{equivclass2}
\ee
$$
J^2_3 ,\quad J^2_1+r^2J^2_2 
,(r\neq \pm 1, |r|\le 1).
$$
We can  proceed as we have in the case of the complex Euclidean 
plane. We  consider one of our two quadratic constants to correspond to one 
of the representatives,(\ref{equivclass2}), hence coming from a
separable coordinate system in standard form. The potential 
must then have an explicit  separable form in the appropriate
coordinates. We then
ask when does there exist an additional quadratic constant and what conditions 
does this impose on our potential. Potentials are considered as
equivalent if they are related by an action of the complex orthogonal
group $O(3)$, including reflections. For background information about
this problem, see \cite{KMJP1,KMJP3}.

Unlike Euclidean space, all superintegrable potentials are multiseparable.
A method for determining the type of separating coordinates from a given
constant is described in Appendix \ref{appendix2} and various possibilities
for each potential found below are listed.

We consider first those systems that separate in horospherical 
coordinates. Thus, there is  a quadratic constant of the form 
$$
A_1={J_-}^2+d_1(x,y,z).
$$
(We adopt the notation $J_\pm=J_1\pm i J_2$ and 
$w=x+iy$, $\bar w=x-iy$.) In terms of
horospherical coordinates $u,v$ this means that the potential can be
represented in the form 
$$
V=f(v)+v^2h(u)
$$
for some functions $f,h$. We now assume that there is
a second constant of the motion. It can be taken in the form
\be
A_2=A{J_+}^2+BJ_3{J_-}+CJ_3{J_+}+DJ_3^2+d_2(u,v).\label{Case1}
\ee
One can show that the case $A\ne 0$ does not admit any nonconstant
potentials. Similarly, the case $A=0,C\ne 0$ doesn't occur. If $A=C=0$
and $D\ne 0$, then via a symmetry transformation $\exp(aJ_-)$, for
suitable $a$ we leave $A_1$ unchanged and map $A_2$ to 
$$
{\tilde A}_2=DJ_3^2+ {\tilde d}_2.
$$
Thus there are only two cases: 1)  $A=C=D=0, B=1$ and 2) $A=B=C=0, D=1$.

\bi
\item[{\bf [S1]}]
$\displaystyle V= {\alpha \over {\bar w}^2} + {\beta z\over {\bar w}^3} + 
{\gamma (1-4z^2)\over {\bar w}^4}$.

\noindent
The extra constant has the form \cite{KMJP3}
$$
A_2=J_3{J_-}+d_2(x,y,z)\,.
$$
\item[{\bf [S2]}]
$\displaystyle V= {\alpha \over z^2} + {\beta \over {\bar w}^2} + 
{\gamma w\over {\bar w}^3}$.

\noindent
The extra constant has the form \cite{KMJP3}
$$
A_2=J^2_3+d_2(x,y,z)\,.
$$
\ei

\noindent
There is a special case of {\bf [S2]} that admits an extra symmetry. 

\bi
\item[{\bf [S3]}]
$\displaystyle V= {\alpha \over z^2}$.

\noindent
The two extra constants are of the form 
$$
A_2=(J_1+iJ_2)^2+d_2(x,y,z),\quad  
A_3=J_3.
$$
For convenience,  we will adopt a modified  basis given by  
$$
{\tilde A}_1=J^2_1+ {\alpha (1+y^2-x^2)\over 2z^2}\,,\quad
{\tilde A}_2=J_1J_2- \alpha  {xy\over z^2}\,,\quad  
X=J_3\,.
$$
The Poisson bracket relations are 
$$
\{X,{\tilde A}_1\}=-2{\tilde A}_2\,,\quad
\{X,{\tilde A}_2\}=-A_0+X^2+2{\tilde A}_1\,,
$$
$$
\{{\tilde A}_1,{\tilde A}_2\}=-X(2{\tilde A}_1+\alpha )\,,
$$
with functional relation 
$$
{\tilde A}_1(A_0-{\tilde A}_1-X^2)-{\tilde A}^2_2- {\alpha \over
2}(X^2+A_0)+\frac{\alpha^2}4 = 0.
$$
\ei

\noindent
We now consider degenerate elliptical coordinates of type 2.  The 
defining constant of the motion has the form 
$$
A_1=J_3J_-+d_1(x,y,z).
$$

There is only one (non-degenerate) new system \cite{KMJP3}

\bi
\item[{\bf [S4]}]
$\displaystyle V = {\alpha \over {\bar w}^2} + {\beta z\over \sqrt{x^2+y^2}} + 
{\gamma \over {\bar w}\sqrt{x^2+y^2}}$.

\noindent
with constant of the motion
$$
A_2=J^2_3+d_2(x,y,z)\,.
$$
\ei

\noindent
There are two special cases of 
{\bf [S4]} that admit an extra constant of the motion:

\bi
\item[{\bf [S5]}]
$\displaystyle V= {\alpha \over {\bar w}^2}$.

\noindent
where the  extra constants can be taken as 
$$
A_1=J_3{J_-}- {\alpha z\over {\bar w}},\quad 
A_2=J^2_3+ \alpha  {w\over {\bar w}},\quad 
X=J_-.
$$
The  Poisson bracket relations take the form 
$$
\{X,A_1\}=iX^2-i\alpha\,,\quad\{X,A_2\}=2iA_1\,,\quad
\{A_1,A_2\}=2iXA_2
$$
with the functional relation 
$$
A^2_1-A_2X^2+\alpha (A_2-A_0)=0\,.
$$

\item[{\bf [S6]}]
$\displaystyle V= {\alpha z\over \sqrt{x^2+y^2}}$.

\noindent
A suitable choice of basis is given by 
$$
A_2=J_1J_3- {\alpha \over 2} {x\over \sqrt {x^2+y^2}}\,,\quad
A_3=J_2J_3- {\alpha \over 2} {y\over \sqrt{x^2+y^2}}\,,\quad 
X=J_3\,.
$$
The Poisson bracket relations are  
$$
\{X,A_2\}=-A_3\,, \quad \{X,A_3\}=A_2\,,\quad 
\{A_2,A_3\}=X(A_0-2X^2)\,,
$$
with the functional relation 
$$
A^2_2+A^2_3+X^4-A_0X^2=0\,.
$$
\ei

\noindent
For degenerate elliptical coordinates of type 1  the constant 
describing this system has the form 
\be
A_1={J_+}^2-J^2_3+d_1(x,y,z).\label{Case3}
\ee
Two new (non-degenerate) potentials arise:

\bi
\item[{\bf [S7]}]
$\displaystyle V= \frac{\alpha x}{\sqrt{ y^2+z^2}} + {\beta y\over z^2\sqrt {y^2+z^2}} + 
{\gamma \over z^2}$.

\noindent
The extra constant has the form 
$$
A_2=J^2_1+d_2(x,y,z),
$$
see \cite{KMJP3}.
\item[{\bf [S8]}]
$\displaystyle V= {\alpha x\over \sqrt {y^2+z^2}} + 
{\beta (w-z)\over \sqrt{ { w}(z-iy)}} + 
{\gamma (w+z)\over \sqrt {w(z+iy)}}$

\noindent
with the second constant given by  
$$
A_2=J_3J_1+d_2(x,y,z).
$$
\ei

\noindent
There are no special  potentials that give a third constant of the
motion in this case. 

We now consider spherical coordinates.  Here, the first constant has the form 
$$
A_1=J^2_3+d_1(x,y,z).
$$

There is one (non-degenerate) new system \cite{KMJP3}. The  potential is

\bi
\item[{\bf [S9]}]
$\displaystyle V= {\alpha \over x^2} + {\beta \over y^2} + {\gamma \over z^2}$.

\noindent
The extra constant has the form 
$$
A_2=J^2_2+d_2(x,y,z).
$$
\ei

\noindent
All of the elliptical coordinate cases have already been covered in
the cases above.
This  completes the list of possible superintegrable potentials 
on the complex two-sphere.

\section{Conclusions}
In this paper we have, in complete generality, enumerated all
potentials on two-dimensional complex constant curvature spaces for 
which there is more than one constant of the motion that is 
quadratic in the momenta.  For each pair of constants of the motion,
whose leading terms are second order in the enveloping algebra of the
Lie symmetry algebra of the free particle Hamiltonian, we find a pair
of coupled second order linear partial differential equations satisfied
by the potential function. The key to making our approach practical is
that when one of the constants of the motion corresponds to a
separable coordinate system, we can explicitly (and simply) solve one
of these PDEs in this coordinate system, and merely have to 
substitute the solution into the second equation.

One can see by inspection of Tables
\ref{Ecoords} and \ref{Scoords} that each of
these cases (except one) is multiseparable, i.e., separation is
possible in at least two coordinate systems. The one counterexample in
flat space ({\bf [E15]})
still separates in one system.  These tables also show that
each potential listed can be uniquely identified by its list of
associated equivalence classes of quadratic constants.  This serves
to confirm that they are indeed distinct potentials, unrelated by
group motions.

We also observe that whenever there is more than one extra 
quadratic constant, a first order constant can be found.  Further, the
non-degenerate potentials found in \cite{KMJP2,KMJP3} that are not
related to a degenerate potential by group motions are those
for which the additional constants are genuinely second order, i.e.\
no first order constant exists.

Note further that for a non-degenerate potential in flat space we can
prescribe $V,V_x,V_y,V_{yy}$ arbitrarily at any regular point $(x_0,y_0)$ and
these values determine $V(x,y)$ uniquely. These potentials correspond
to exactly three functionally and linearly independent constants of
the motion. For a degenerate potential with  an extra (linearly
independent) constant of the motion the additional constant implies a
relationship between $V_x,V_y$ at any regular point; hence that all 
first, second and higher order derivatives of $V(x,y)$ can be
expressed in terms of a single first derivative, say $V_x$. Thus for
all these potentials we can prescribe $V,V_x$ arbitrarily at any
regular point  and
these values determine $V(x,y)$ uniquely. It follows that except for
the exceptional case ({\bf [E15]}) the superintegrable potentials
depend on exactly 4 or 2 parameters. Analogous comments hold for the
complex two-sphere, except that here there is no exceptional case.

What is exceptional about ({\bf [E15]})? This is the only case where one cannot
solve for $V_{xx}-V_{yy}$ and $V_{xy}$ as linear combinations of
$V_x,V_y$. Thus the potential must be degenerate. Indeed this
potential, although it depends on an infinite number of parameters, must
have the form $V({\bar z})$ so $V_x$ and $V_y$ cannot be prescribed
independently at a point. Furthermore, the potential is not uniquely
determined by the values of $V,V_x,V_y,V_{yy}$ at a point.

We give in this paper, and
preceding papers, the structure of the classical quadratic algebras
in almost all cases. We intend to perform  a comprehensive study of
the corresponding  quantum algebras associated with the Schr\"odinger
equation at a later date.

\begin{appendix}

\section{Separable coordinates in $E_{2,C}$}
\label{appendix1}

Each coordinate system in which the Hamilton-Jacobi equation 
is separable on 
$E_{2,C}$ is characterized by a constant quadratic in the momenta.
Coordinate systems that are related by Euclidean group motions
belong to the same family and hence a given family of
coordinates (e.g.\ polar coordinates) 
is associated with an equivalence class of
quadratic elements in the enveloping algebra of $e(2,C)$.  Two elements
are equivalent if one can be transformed into to other by a combination of
scalar multiplication, addition of multiples of $p_x^2+p_y^2$ and
Euclidean motions (including reflections).  One equivalence class 
(listed below) is not associated with a separating coordinate system.
The following can be taken as a representative list of
coordinate systems and corresponding constants.

\begin{enumerate}
\item Cartesian coordinates.  
$$
x\,,y\,, \qquad     L=p^2_x\,.
$$
\item Light cone coordinates.
$$
z = x+iy\,, \quad \bar z = x-iy\,, \qquad L = (p_x+ip_y)^2\,.
$$
\item Polar coordinates.  
$$
x_S=r\cos\theta\,,  \quad
y_S=r\sin\theta\,,  \qquad     L=M^2\,.
$$
\item Semi-Hyperbolic coordinates.
$$
x_{SH}=i(w-u)^2+2i(w+u)\,, \quad 
y_{SH}=-(w-u)^2+2(w+u)\,,
$$
$$
L=M(p_x+ip_y)+(p_x-ip_y)^2\,.
$$
\item Hyperbolic coordinates.  
$$
x_H= {r^2+s^2+r^2s^2\over 2rs}\,,\quad y_H=i {r^2+s^2-r^2s^2\over 2rs}\,,\quad 
L=M^2+(p_x+ip_y)^2\,.
$$
\item Parabolic coordinates.  
$$
x_P=\xi \eta\,,
\quad 
y_P={1\over 2}(\xi ^2-\eta ^2)\,,
\qquad 
L=Mp_x\,.
$$
\item Elliptic coordinates.
$$
x_E=c\sqrt{(u-1)(v-1)}\,,\quad y_E=c\sqrt {-uv}\,,\qquad 
L=M^2+c^2p^2_x\,.
$$
\item No separation.
$$
\mbox{No corresponding separable coordinates,} \qquad
L = M(p_x + ip_y)\,.
$$
\end{enumerate}

The following facts are useful in determining to which class given 
constant belongs.

\begin{itemize}
\item
Translations leave $p_x$ and $p_y$ unchanged and
for any $A$ and $B$ a translation can be found that has the effect
$$
M\to M + Ap_x + Bp_y\,.
$$
\item
Rotations leave $M$ fixed and one can be found that for any $A$ and
$B$ has the effect
$$
Ap_x+Bp_y \to \mbox{one of $p_x$, $p_+$ or $p_-$.}
$$
\item
A rotation can be found that for any $A\neq0$ has the effect
$$
p_+\to Ap_+\,, \quad \mbox{and} \quad p_-\to\frac1Ap_-\,.
$$
\item
The reflection $y\to -y$ has the effect
$$
M\to -M\,, \qquad p_+ \longleftrightarrow p_-\,.
$$
\end{itemize}

For each superintegrable potential {\bf [E1--20]}, all linear
combinations of the given quadratic constants must be considered in order
to determine which equivalence 
classes are represented, and hence in which families
of coordinates systems it will separate.

For example, the potential {\bf [E18]} has constants with leading part
$L=AM^2+BMp_x+CMp_y+D(p_x^2+p_y^2)$.  From this we can see immediately
that polar and parabolic coordinates will separate this Hamiltonian, and
that the non-separating constant $Mp_+$ can be generated.
Further, $M^2+2Mp_y+2(p_x^2+p_y^2)\to M^2+p_x^2$ under a translation that
maps $M\to M-p_y$ and hence {\bf [E18]} separates in
an elliptic coordinate system.  Lastly, there exists a translation mapping
$M^2+2iMp_+\to M^2+p_+^2$, and 
hence the Hamiltonian separates
in hyperbolic coordinates.

The results of similar reasoning for the other potentials are summarized
in table \ref{Ecoords}.

\section{Separable coordinates in $S_{2,C}$}
\label{appendix2}

As for Euclidean space, coordinates separating the Hamilton-Jacobi equation
on the two-sphere correspond to constants that are 
quadratic in the elements of 
the Lie algebra of its symmetry group $O(3)$.  Coordinates belong to 
the same family if one can be transformed to the other by
a rotation or reflection.  On the complex two-sphere, unlike 
complex Euclidean space,
every quadratic constant, other than a multiple of the Hamiltonian, 
corresponds to a separating coordinate system.  

The separable coordinates on the complex two-sphere and their 
characterizing constants are:

\begin{enumerate}
\item Spherical coordinates. 
$$
x=\sin\theta\cos\varphi\,,\quad y=\sin\theta \sin\varphi\,,
$$
$$ 
z=\cos\theta\,, \qquad  L=J^2_3\,.
$$
\item Horospherical coordinates.
$$
x=\frac i2\left(v + \frac{u^2-1}v\right)\,, \quad 
y=\frac12\left(v + \frac{u^2+1}v\right)\,,
$$
$$  
z=\frac{iu}v\,, \qquad L=(J_1-iJ_2)^2\,.
$$
\item Elliptic coordinates.
$$
x^2= {(ru-1)(rv-1)\over 1-r}\,,\quad y^2= {r(u-1)(v-1)\over r-1}\,,
$$
$$  
z^2=ruv\,, \qquad  L=J^2_1+rJ^2_2\,.
$$
\item Degenerate Elliptic coordinates of type 1.
$$
x+iy = {4uv\over (u^2+1)(v^2+1)}\,,\quad x-iy = 
{(u^2v^2+1)(u^2+v^2)\over uv (u^2+1)(v^2+1)}\,,
$$
$$  
z = {(u^2-1)(v^2-1)\over (u^2+1)(v^2+1)}\,, \qquad  L=(J_1+iJ_2)^2-J^2_3\,.
$$
\item Degenerate Elliptic coordinates of type 2.
$$
x+iy=-iuv\,,\quad x-iy={1\over 4} {(u^2+v^2)^2\over u^3v^3}\,,
$$
$$ 
z=\frac i2\frac{u^2-v^2}{uv}\,, \qquad  L=J_3(J_1-iJ_2)\,.
$$
\end{enumerate}

The action of the symmetry group on a general quadratic constant is not
as easily described as for $E_{2,C}$.  
To determine the equivalence class to which 
a given quadratic element $L$ belongs it is more convenient to note that 
the number of distinct eigenvalues of $L$, as a quadratic form in the
$J_i$, and the dimension
of the kernel of the map on first order elements, 
$\phi: X \mapsto \{X,L\}$, are both invariant under group motions and
addition of multiples of the Casimir $J_1^2+J_2^2+J_3^2$.
Table \ref{Sinvariants} gives the correspondence between these
invariants and families of coordinate systems on $S_{2,C}$.

Just as for $E_{2,C}$, by considing a general linear combination of
constants for each potential {\bf [S1--S9]}, the corresponding families
of separable coordinates can be determined.  The results are summarized
in table \ref{Scoords}.

\begin{table}
\begin{tabular}{r||c|c|}
  & no.\ of distinct eigenvalues & $\dim\ker(\phi)$ \\
\hline
\hline
Spherical                  & 2 & 1 \\
\hline
Horospherical              & 1 & 1 \\
\hline
Elliptic                   & 3 & 0 \\
\hline
Degenerate elliptic type 1 & 2 & 0 \\
\hline
Degenerate elliptic type 2 & 1 & 0 \\
\hline
\end{tabular}
\caption{Invariants used to identify coordinate systems on $S_{2,C}$}
\label{Sinvariants}
\end{table}

\begin{table}
\begin{tabular}{r|*{10}{|c}|}
\makebox[26mm]{}&\bb{E1}  &\bb{E2}  &\bb{E3}  &\bb{E4}  &\bb{E5}
                &\bb{E6}  &\bb{E7}  &\bb{E8}  &\bb{E9}  &\bb{E10}\\
\hline\hline
%                 E1   E2   E3   E4   E5   E6   E7   E8   E9   E10
Cartesian       & \x & \x & \x & \x & \x & \x &    &    &    &    \\
\hline
Light Cone      &    &    & \x & \x & \x &    & \x & \x & \x & \x \\
\hline
Polar           & \x &    & \x &    &    & \x &    & \x &    &    \\
\hline
Semi-Hyperbolic &    &    &    & \x &    &    &    &    &    & \x \\
\hline
Hyperbolic      &    &    & \x &    &    &    & \x & \x &    &    \\
\hline
Parabolic       &    & \x &    &    & \x & \x &    &    & \x &    \\
\hline
Elliptic        & \x &    & \x &    &    & \x & \x &    &    &    \\
\hline
Non-Separating  &    &    &    & \x &    &    &    &    &    &    \\
\hline
\end{tabular}

\vskip1cm

\begin{tabular}{r|*{10}{|c}|}
\makebox[26mm]{}&\bb{E11} &\bb{E12} &\bb{E13} &\bb{E14} &\bb{E15}
                &\bb{E16} &\bb{E17} &\bb{E18} &\bb{E19} &\bb{E20} \\
\hline\hline
%                E11  E12  E13  E14  E15  E16  E17  E18  E19  E20
Cartesian       &    &    &    &    &    &    &    &    &    &    \\
\hline
Light Cone      & \x & \x & \x & \x & \x &    &    &    &    &    \\
\hline
Polar           &    &    &    & \x &    & \x & \x & \x &    &    \\
\hline
Semi-Hyperbolic & \x &    & \x &    &    &    &    &    &    &    \\
\hline
Hyperbolic      &    & \x &    & \x &    &    & \x & \x & \x &    \\
\hline
Parabolic       &    &    & \x &    &    & \x &    & \x &    & \makebox[12pt]{\x\x}   \\
\hline
Elliptic        &    & \x &    &    &    & \x &    & \x & \x &    \\
\hline
Non-Separating  & \x & \x & \x & \x & \x &    & \x & \x & \x & \x \\
\hline 
\end{tabular}
\caption{Separating coordinate systems for superintegrable potentials
in complex two-dimensional Euclidean space.  Potentials possessing
a quadratic constant equivalent to $Mp_+$ are indicated in
the line labelled `Non-Separating'.}
\label{Ecoords}
\end{table}

\begin{table}
\begin{tabular}{r|*{9}{|c}|}
\makebox[26mm]{}&\bb{S1} &\bb{S2} &\bb{S3} &\bb{S4} &\bb{S5}
                &\bb{S6} &\bb{S7} &\bb{S8} &\bb{S9} \\
\hline\hline
%                       S1   S2   S3   S4   S5   S6   S7   S8   S9
Spherical             &    & \x & \x & \x & \x & \x & \x &    & \x \\
\hline
Horospherical         & \x & \x & \x &    & \x &    &    &    &    \\
\hline
Elliptic              & \x &    & \x &    & \x & \x & \x & \x & \x \\
\hline
Degenerate elliptic 1 &    & \x & \x & \x & \x & \x & \x & \x &    \\
\hline
Degenerate elliptic 2 & \x &    &    & \x & \x & \x &    &    &    \\
\hline
\end{tabular}
\caption{Separating coordinate systems for superintegrable potentials
on the two-dimensional complex sphere.}
\label{Scoords}
\end{table}

\end{appendix}


\begin{thebibliography}{99}
%-----------------------------------------------------------------------
\bibitem{VIA} V.I. Arnold.
Mathematical Methods of Classical Mechanics.
(translated by K. Vogtmann and A. Weinstein)
{\it Graduate Texts in Mathematics, 60,
 Springer-Verlag},
 New York, 1978.
%-----------------------------------------------------------------------
\bibitem{EVA}N.W.Evans. 
Superintegrability in Classical Mechanics;
{\it
Phys.Rev.}\ {\bf A 41} (1990) 5666; Group Theory of the
Smorodinsky-Winternitz System; {\it J.Math.Phys.}\ {\bf 32},  3369
(1991).
%-----------------------------------------------------------------------
\bibitem{EVAN}
N.W.Evans.
Super-Integrability of the Winternitz System;
{\it Phys.Lett.}\ {\bf A 147},  483 (1990).
%-----------------------------------------------------------------------
\bibitem{WOJ}
S.Wojciechowski.
Superintegrability of the Calogero-Moser System.
{\it Phys. Lett.} {\bf A 95},  279 (1983);
%-----------------------------------------------------------------------
\bibitem{RAN}  M.F. Ra\~nada.
Superintegrable $n$=2 systems, quadratic constants of motion, and
potentials of Drach.
{\it J.Math.Phys.} {\bf 38}, 4165, (1997).
%-----------------------------------------------------------------------
\bibitem{KMJP2}  E.G.Kalnins, W.Miller Jr. and G.S.Pogosyan.
Completeness of multiseparable superintegrability in $E_{2,C}$.
 {\it J.Phys.A: Math Gen.} {\bf 33}, 4105, (2000).
%-----------------------------------------------------------------------
\bibitem{KMJP3}  E.G.Kalnins, W.Miller Jr. and G.S.Pogosyan.
Completeness of multiseparable superintegrability on the complex
2-sphere,   {\it J.Phys.A: Math Gen.} {\bf 33}, 6791, (2000).
%-----------------------------------------------------------------------
\bibitem{KMJP4}  E.G.Kalnins, W.Miller Jr. and G.S.Pogosyan.
Completeness of multiseparable superintegrability in two dimensions.
To appear in {\it Proceedings of the XXIII International 
Colloquium on Group Theoretical Methods in Physics, Dubna, Russia,
July 31 -- August 5, 2000}.
%-----------------------------------------------------------------------
\bibitem{MIL}W.Miller, Jr.
Symmetry and Separation of Variables.
{\it Addison-Wesley Publishing Company},\ Providence, Rhode Island, 1977.
%-----------------------------------------------------------------------
\bibitem{KMJP1}  E.G.Kalnins, W.Miller Jr. and G.S.Pogosyan.
Superintegrability and associated polynomial solutions: Euclidean space and
the sphere in two dimensions.
{\it J.Math.Phys.} {\bf 37}, 6439, (1996).
%-----------------------------------------------------------------------
\end{thebibliography}
\end{document}